\documentstyle[12pt]{article}
\input epsf
\title{Tests for Cosmological Evolution of a Brane Universe Model
}

\author{Xinhe MENG (1,2,3) \footnote{
mengxh@public.tpt.tj.cn}
,
 \ \ Jingmin HOU (1), Kaiyi LU (1) and Wenyao ZHAO (1)
\\ 1  Department of Physics, Nankai University, Tianjin 300071, P.R.China\\
2 Institute of Theoretical Physics, CAS, Beijing 100080, P.R.China\\
3 Department of Physics, Hiroshima University, Hiroshima 739-8521, Japan}
\begin{document}

\maketitle

\begin{abstract}
The relativistic Friedmann Lemaitre cosmology model (FLCM) is very
sucessful to describe the evolution history of the Universe from
the " First three Minutes". Any alternative model should be
consistent with  the FLCM explanations to the later stage
evolutions of the Universe at certain points. An noncompact extra
dimension model was recently proposed by Randall and Sundrum.
Binetruy et al. obtained the modified Friedmann equation, in
which the energy density of the brane appears quadratically in
contrast with the linear behavior of the standard Friedmann
equation.  We investigate kinds of classical cosmological effects of
the new models and get a general solution of the cosmic evolution
for this extended model, with more detail discussions of the brane
tension parameter on these cosmological tests.


\end{abstract}

\section{Introduction}
Brane world cosmology seems to provide an alternative explanation
for the present accelerating stage of the Universe\cite{pr} without
introducing either a cosmological constant or an evolutional
quintessence-like component. The basic idea in these scenarios is
the existence of a higher dimensional bulk in which our Universe
is sitting as a hypersurface, named 3-brane. The particle physics standard
model matter fields are located in the brane while gravity
propogates in all dimensions, namely bulk. The large extra
dimensions also can solve the mass hierarchy problem of the
standard particle physics model. Previously it has been assumed
that the extra dimensions are compact to be unobservable small, as
the  so-called Kaluza-Klein model\cite{kaluza,klein}. Recently
Randall and Sundrum relaxed the constraints of Kaluza-Klein model
and proposed a new idea \cite{rs1,rs2} that
 the extra dimension is noncompact, which was also proposed earlier by Wesson\cite{wes} (and for the equivalence between them see for example, J.Leon,Mod.Phys.Lett.A16 (2001) 2291), i.e., the extra dimension  is large and
conceivable.

The relativistic Friedmann Lemaitre cosmology model (FLCM) is very
successful to describe the evolution history of the Universe from
the " First three Minutes". Any alternative models should match
the FLCM explanations to the later stage evolutions of the
Universe at certain points. How to test the  Randall-Sundrum
model? The best way is to investigate its cosmological effects,
which can provide constraints on these type of models. Binetruy et
al. have researched the cosmological evolution of the Universe and
shown that the equations governing the cosmological evolution of
the brane will be different from the analogous Friedmann equations
of standard cosmology \cite{binetruy}, i.e., the
 energy density
of the brane Cosmology appears quadratically in the new Friedmann
equation in contrast with the linear behavior of the usual
equation. To reconcile brane cosmology with the standard
cosmological scenario, they put a cosmological constant in the
bulk and introduce in the brane, in addition to ordinary matter
fields, a constant tension that exactly compensates the bulk
cosmological constant, so that the quadratic term would be
cancelled, to leave the linear term dominated in low energy
evolution stages as argued. In this paper we mainly discuss the
additional quadratic brane density term of the brane cosmology
model  which is obviously different from  the standard Friedmann
one, trying to give more qualitative relations to see the effects
from the unconstrained tension parameter, with the hope to give
more physics limits to the parameter.

First, we briefly review the solutions presented by Binetruy et
al\cite{binetruy},  from their generalized Friedmann equation,
\begin{equation}
\frac{\dot{a}^2_0}{a^2_0}=\frac{\kappa^4}{36}\rho^2_b+\frac{C}{a^4_0}-\frac{k}{a^2_0}+\frac{\kappa^2}{6}\rho_B
\label{aa}
\end{equation}
where $a_0$   is scale factor in brane when the coordinate of the
fifth dimension $y=0$. The constant $\kappa$  is related to the
five-dimensional Newton's constant $G_{(5)}$   and the
five-dimensional reduced Planck mass $M_{(5)}$  by the relations
$\kappa^2=8\pi G_{(5)}=M^{-3}_{(5)}$. The $\rho_B$ and $\rho_b$
are the energy density in bulk and in brane respectively. The $C$
is an integral constant that may be interpreted as a dark
radiation term for the sake of radiation dominated behavior form,
and $k$ is the spatial curvature. From Eq.(\ref{aa}), we know that
square of Hubble parameter $H^2=\dot{a}^2_0/a^2_0$ is quadric of
the energy density in brane $\rho_b$ , while it is linear to
$\rho_b$ in the
 standard cosmology.
So it must be able to be reduced to the standard theory, i.e. linear to $\rho_b$,
if it is reasonable.
Binetruy, et al. assumed that the energy density
 in the brane could be decomposed in two parts,
\begin{equation}
\rho_b=\rho_\Lambda+\rho
\label{ab}
\end{equation}
where $\rho_\Lambda$  is a constant that represents an intrinsic
tension of the brane and $\rho$ stands for the ordinary energy
density in cosmology. Substituting in Eq.(\ref{aa}) one gets
\begin{equation}
\frac{\dot{a}^2_0}{a^2_0}=\frac{\kappa^2}{6}\rho_B+\frac{\kappa^4}{36}\rho^2_\Lambda
+\frac{\kappa^4}{18}\rho_\Lambda\rho_b+\frac{\kappa^4}{36}\rho^2_b
+\frac{C}{a^4_0}-\frac{k}{a^2_0} \label{ac}
\end{equation}
Following Randall and Sundrum\cite{rs2}, Binetruy, et al. by
choosing $\rho_\Lambda$   such that
\begin{equation}
\frac{\kappa^2}{6}\rho_B+\frac{\kappa^4}{36}\rho^2_\Lambda=0
\label{ad}
\end{equation}
and we then get
\begin{equation}
\frac{\dot{a}^2_0}{a^2_0}=\frac{8\pi
G}{3}(\rho+\frac{\rho^2}{2\lambda})
+\frac{C}{a^4_0}-\frac{k}{a^2_0}
\label{ae}
\end{equation}
where it is assumed that $8\pi G=\frac{\kappa^4\rho_\Lambda}{6}$
and $\lambda=1/\rho_\Lambda$. If the brane tension parameter
$\lambda\rightarrow\infty$ and the integral constant
$C=0$,Eq.(\ref{ae}) will be reduced to the standard Friedmann
equation
\begin{equation}
\frac{\dot{a}^2_0}{a^2_0}=\frac{8\pi G}{3}\rho-\frac{k}{a^2_0}
\label{af}
\end{equation}
Throughout this paper we neglect the possibility of the existence of a cosmological constant on the brane for the early universe evolution and for simplicity\cite{se}, to focus on the brane tension effects. For the inclusion of a cosmological constant or a quintessence-like scalar field with negative equation of state is now under our consideration in another work to appear. 

We have in addition the usual equation of conservation for the energy-momentum
tensor of the cosmic fluid given by
\begin{equation}
\dot{\rho}+3H(p+\rho)=0
\label{ag}
\end{equation}

In this paper, we assume that the integral constant $C$ in
Eq.(\ref{ae}) is equal to zero for simplicity, or for the later
evolutions of the Universe the dark radiation term negligible, and
$k=0$ as the recent observational data suggest that the Universe
is flat\cite{pryke, netterfield, stompor}. In this framework we
mainly discuss several classical observational tests for this
enlarged brane cosmology model to see the effects of the
additional term.

\section{The critical density}
It is convenient to discuss dimensionless quantity as we often do
in physics investigation. In the standard cosmology, for
convenience, the density fraction of the world models is expressed
in terms of a critical density, which is defined to be the density
when the curvature $k=0$, i.e.,  $\rho^{standard}_c=3H^2/8\pi G$.It
is similar  that we can define the critical density in the brane
Universe models. By choosing the integral constant $C=0$ in
Eq.(\ref{ae}), we get
\begin{equation}
H^2=\frac{\dot{a}^2}{a^2}=\frac{8\pi
G}{3}(\rho+\frac{\rho^2}{2\lambda})-\frac{k}{a^2}
\label{ba}
\end{equation}
where, for convenience, we have omitted the subscript of the scale
factor, which would not lead to confusion in this paper. By
assuming the curvature $k=0$ in Eq.(\ref{af}), we deduced the
critical density
\begin{equation}
\rho_c=\sqrt{\lambda^2+\frac{3H^2\lambda}{4\pi G}}-\lambda
\label{bb}
\end{equation}

\begin{figure}[ht]
\epsfxsize=14cm
\epsfbox{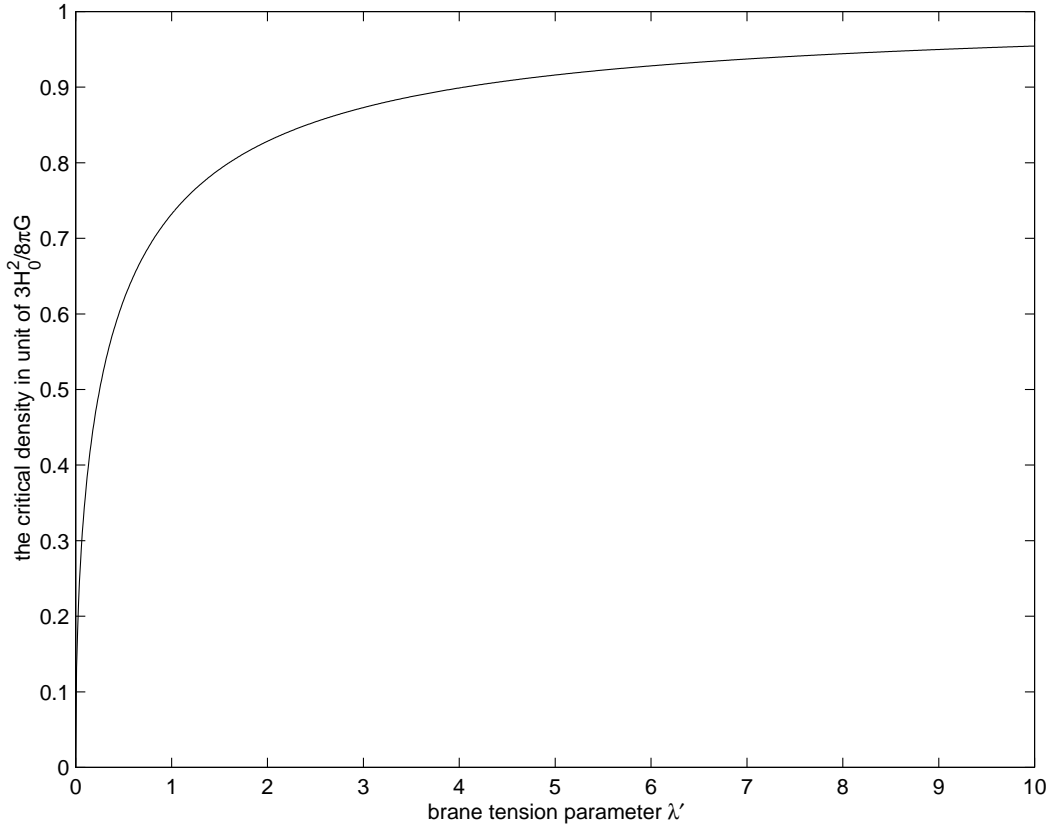}
\caption{\small The critical density in the brane Universe as a function
of brane tension parameter. The brane tension parameter and the
critical density are both in units of
$\rho^{standard}_c=3H^2/8\pi G$}
\label{density}
\end{figure}
For convenience, we define the dimensionless brane tension
parameter $\lambda '=8\pi G\lambda /3H^2_0$. Note that the
critical $\rho_c$ is  a monotone increasing function of brane
tension parameter $\lambda '$, and that when
$\lambda\rightarrow\infty$, $\rho_c$ reduces to
$\rho_c^{standard}$, as we expect. More details of the
relationship between the critical density $\rho_c$ and brane
tension parameter are shown in Figure \ref{density}. The critical
density sharply  increases with $\lambda$ in the lower values of
$\lambda '$, and the curve becomes flat when $\lambda '>2$. When
$\lambda '$ is in larger values,in the Eq.(\ref{ba}), the linear
term is the leading term, and the quadratic term gives small
correction to the standard Friedmann equation; when $\lambda$ is
in smaller values, the quadratic term becomes dominated, which
indicates that the brane tension plays an important role in the
energy density of the brane cosmology model.

\section{A general evolution solution of the brane Universe}
In the standard cosmology, one usually uses the comoving
coordinates to describe the dynamical evolution states, in which
the coordinates of any place don't vary with time, and the
evolution of the Universe is described by a scale factor that is a
function of cosmic time. If one knows the relationship between the
scale factor and the cosmic time, then the global dynamical
behavior about the evolution of the Universe is clear. For
example, one can directly deduce the Hubble parameter and the
deceleration parameters as functions of time, and the dependence
relation of the luminosity distance on redshift, and so on. So for
a cosmological model to get its dynamical evolution solution, we
will be able to obtain kinds of observable quantities to be
compared with the experimental data for the model testings.

\begin{figure}[ht]
\epsfxsize=14cm
\epsfbox{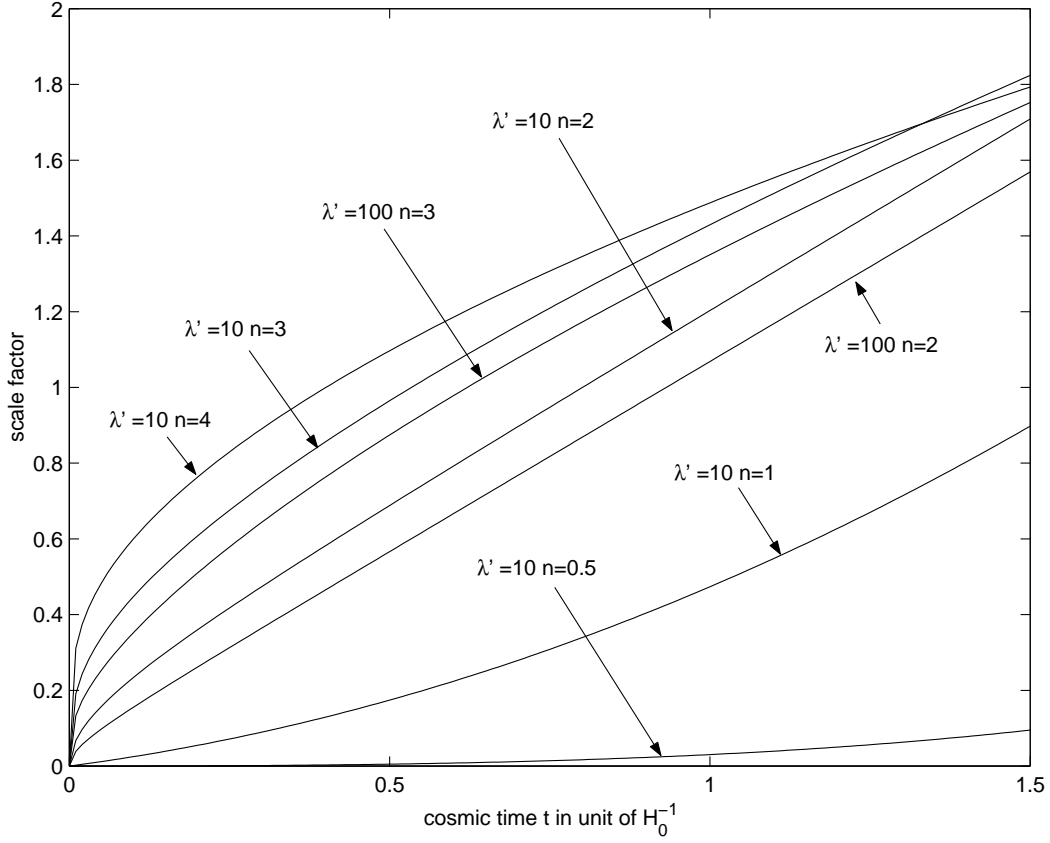}
\caption{\small The scale factor as a function of cosmic time}
\label{scale}
\end{figure}

 In this section we will consider a general case, where the state equation
of matter and energy is
\begin{equation}
p=w\rho
\label{ea}
\end{equation}
and $w$ is assumed to be a constant for simplicity in this paper.
From Eqs.(\ref{ag}) and (\ref{ea}) ,  one gets
\begin{equation}
\rho=\rho_0a^{-n}
\label{eb}
\end{equation}
where $\rho_0$  is a constant standing for the energy density of
the Universe today, $n$ is a parameter related to the equation of
state parameter w by $n=3(1+w)$. For recent observational data
strongly suggest that the Universe is flat, i.e. $k=0$, so we only
investigate the case when $k=0$ . Substituting Eq.(\ref{eb}) in Eq.
(8) and taking $k=0$, we get
\begin{equation}
H^2=H^2_0\Omega_0a^{-n}+\frac{H^2_0\Omega^2_0}{2\lambda '}a^{-2n}
\label{ec}
\end{equation}
where $H_0$  is the Hubble parameter today, $\Omega_0$  and $\lambda '$ are defined
by $\Omega_0=3H^2_0\rho_0/8\pi G$  and $\lambda '=8\pi G\lambda /3H^2_0$  respectively.
Assuming the initial condition $a(t)|_{t=0}=0$, we get a general
solution of Eq.(\ref{ec})
\begin{equation}
\begin{array}{ll}
a(t)=\left [\left (
\frac{\sqrt{n^2H^2_0\Omega_0}}{2}t+\sqrt{\frac{\Omega_0}{2\lambda
'}}\right)^2-\frac{\Omega_0}{2\lambda '}\right]^{1/n} & n>0\\
a(t)\propto \exp \left(\sqrt{\frac{2\lambda
'H^2_0\Omega_0}{2\lambda '+\Omega_0}}t \right) & n=0
\end{array}
\label{ee}
\end{equation}
The last solution is consistent with the inflation explanation if
the energy scale is very high or we may use it to describe the
Universe later accelerating expanding stage without a cosmological constant when the energy scale
is very low, as observed today. Assuming $\Omega_0=1$, we get the
relationship between the scale factor and the cosmic time with
different parameters respectively, which are shown in Figure
\ref{scale}. Comparing the two curves about $n=2$, we find that
for the brane tension parameter smaller, the scale factor $a$ will
increase faster, and for the case of $n=3$ we get the same
conclusion. In another way, the curves with larger state equation
parameter $n$ grow faster in one Hubble time scale
 in contrast to ones with smaller parameter $n$ when we fix the value of $\lambda '$, for example 10.
\section{The deceleration parameter}
Following the fact that our Universe is expanding, one may ask a
question that the expansion of the Universe is accelerating or
decelerating, or how fast it expands. So, besides the Hubble
parameter, the deceleration parameter is another important
parameter describing the evolution of the Universe to measure the
rate of slowing of the expansion , which is defined by
\begin{equation}
q=-\frac{a\ddot{a}}{\dot{a}^2}
\label{fa}
\end{equation}
where $q$ is dimensionless. Substituting Eq.(\ref{ec})and its
derivative in Eq.(\ref{fa}), one gets
\begin{equation}
q=\frac{(n-2)\lambda 'a^n+(n-1)\Omega_0}{2\lambda 'a^n+\Omega_0}
\label{fb}
\end{equation}

In the standard cosmology, the deceleration parameter $q$ is a
function of the present density parameter $\Omega_0$ and the state
equation parameter $n$. It is here also a function of the brane
tension parameter $\lambda '$ in the brane cosmology, which we can
see from the above equation. In the present epoch the scale factor
can be set $a=1$, so we can get the deceleration parameter today,
\begin{equation}
q_0=\frac{(n-2)\lambda '+(n-1)\Omega_0}{2\lambda '+\Omega_0}
\label{fc}
\end{equation}

\begin{figure}[ht]
\epsfxsize=14cm \epsfbox{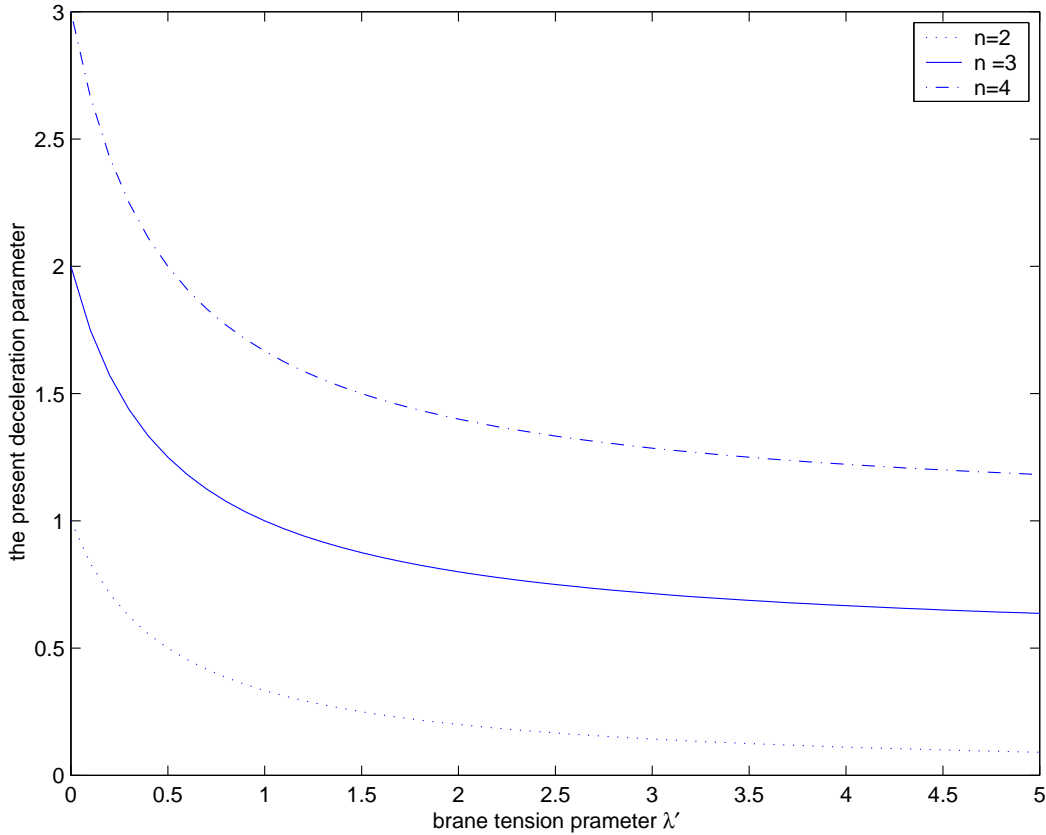} \caption{\small The present
deceleration parameter as a function of brane tension parameter
$\lambda$} \label{dec1}
\end{figure}

Figure \ref{dec1} shows that the relationship between the present
deceleration parameter and the  brane tension parameter $\lambda
'$ for $n=2,3,4,$ respectively. The present deceleration parameter
is a monotone function of the brane tension parameter.The present
deceleration parameter sharply falls for lower values of the brane
tension parameter, and the curves become flat when $\lambda '>2$.

From mainly the recent SN Ia and Cosmic Microwave Background observations\cite{pr} a cosmological constant-like dark energy may exist today which powers the Universe expansion accelerating with the decelerating parameter -1< q <0. Therefore this observational constraint is not satisfied in the present model without a cosmological constant-like term. Certainly we need more SN Ia observational data to confirm the important constraint that is just the main mission of the near future SNAP project. When the brane parameter approches infinite large, as in the low energy scale case in our present discussion, the decelerating parameter goes to the standard cosmological one without the dark energy existence. The brane world cosmology model with a cosmological constant-like term for a possible negative decelerating parameter is under our investigation in the paper to appear\cite{mx}, but the analytic relations are more complicated than the present ones
. 
\section{The cosmic time-redshift relation and the age of brane
Universe}

\begin{figure}[ht]
\epsfxsize=14cm
\epsfbox{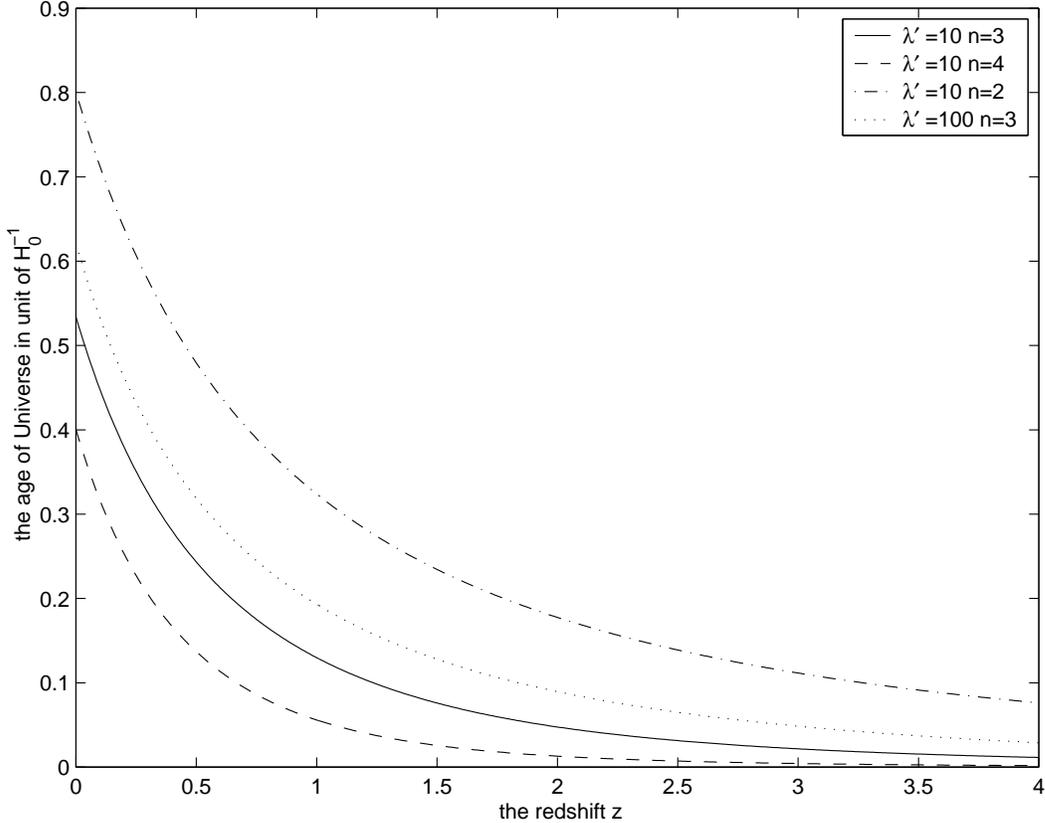}
\caption{\small The age of Universe as a function of  redshift
$z$}
\label{age}
\end{figure}

An important result for many of cosmology models is the relation
between cosmic time $t$ and redshift $z$, from which one can
calculate the age of the Universe. So one can compare the
calculated age of the Universe and the age of the oldest stars in
globular clusters to give constraints on the cosmological models.
Because $a=(1+z)^{-1}$, it follows immediately from Eq.(\ref{ec})
that
\begin{equation}
\frac{dz}{dt}=-(1+z)\left
[H^2_0\Omega_0(1+z)^n+\frac{H^4_0\Omega^2_0}{2\lambda
'}(1+z)^{2n}\right ]^{1/2}
\label{ga}
\end{equation}
Cosmic time $t$ measured from the Big Bang following by the
integration
\begin{equation}
t(z)=\int_0^t
dt=-\frac{1}{H_0}\int_\infty^z\frac{dx}{(1+x)\left[\Omega_0(1+x)^n+
\displaystyle\frac{\Omega^2_0}{2\lambda
'}(1+x)^{2n}\right ]^{1/2}}
\label{gb}
\end{equation}
This integral is difficult to evaluate, so we only work out the
numerical solutions for different values of the brane tension
parameter, which are shown in Figure \ref{age}. The cosmic time
$t$ is a decreasing function of the redshift $z$. From the curves
for $\lambda '=10$, we can see that larger state equation
parameter implies larger age of the Universe; from the two curves
for fixed $n=3, \lambda '=10$ and 100 respectively, we know that
larger brane tension parameter also implies larger age of the
Universe.

\section{The dependence of luminosity distance on redshift }
 Hubble parameter is one of the best parameters describing the
 evolution of Universe, to express the expanding rate, but it can't be measured directly.  We
 have to measure it indirectly, i.e., to deduce it from other parameters
 which can be measured. In the standard cosmology, one gets its value from Hubble's
 law, which
 is the linear relationship between the ``distance'' to a galaxy and its observed
 redshift. So if we know the relationship between the ``distance''
 and the redshift, we will know the value of the Hubble parameter.
 Usually, one uses the luminosity distance as the ``distance'' to
 a galaxy, which is defined by $d^2_L\equiv {\cal L} / 4 \pi
\cal F$, where $\cal L$ is the absolute luminosity of a galaxy and $\cal F$ is
 the measured energy flux, i.e., the energy per time per area
 measured by a detector.

\begin{figure}[ht]
\epsfxsize=14cm \epsfbox{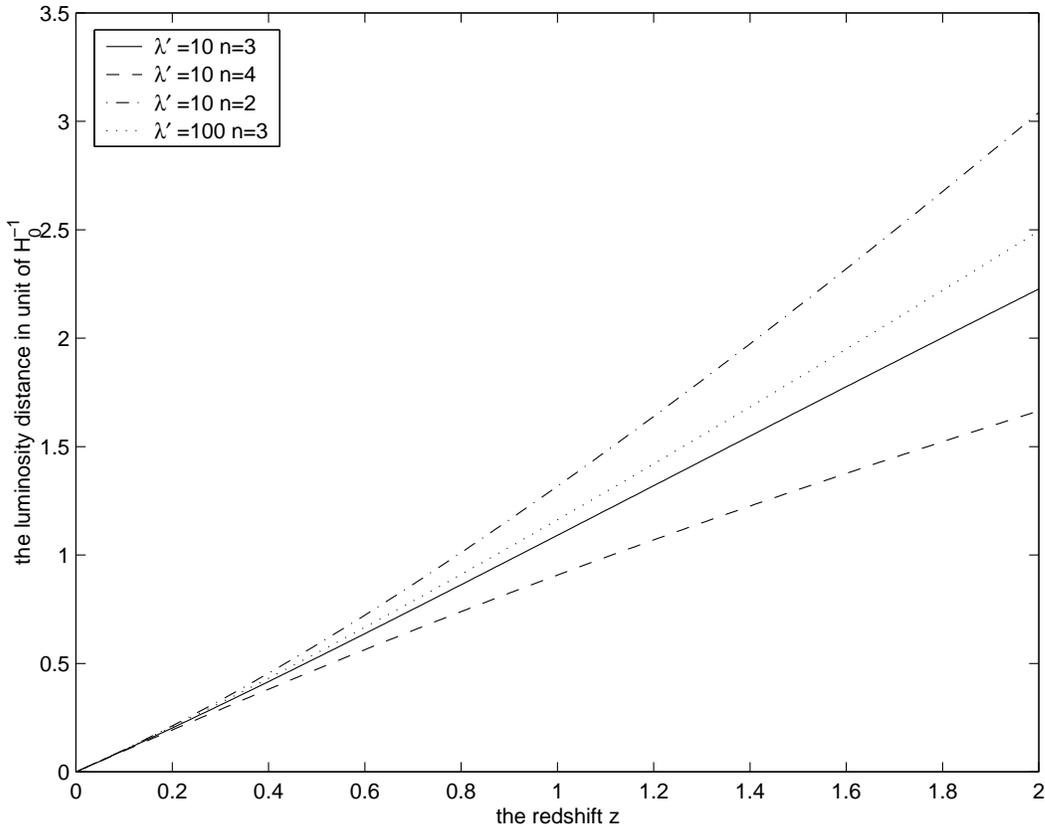} \caption{\small The luminosity
distance as a function  of redshift $z$} \label{distance}
\end{figure}

 In order to compare the results of the model with 
 cosmological tests we must deduce the relationship between the
 luminosity  distance and the redshift from other relations, see Refs
 \cite{dis, dist}. The deduced relationship between the
 luminosity distance and the redshift from the relationship between Hubble
 parameter and the redshift,  as follows
\begin{equation}
 d_L=(1+z)\left \{
    \begin{array}{ll}
    \sin (\sqrt{|\Omega_k|}d_C)/\sqrt{|\Omega_k|} & k=1\\
    d_C &  k=0\\
    \sinh (\sqrt{|\Omega_k|}d_C)/\sqrt{|\Omega_k|} & k=-1
    \end{array}
    \right.
\label{ha}
\end{equation}
where $d_C$ is defined by
\begin{equation}
d_C=\int_0^z H_0\frac{dx}{H(x)}
\label{hb}
\end{equation}
Substituting $a=(1+z)^{-1}$ in Eq.(\ref{ec}), we get
\begin{equation}
H(z)=H_0\left [\Omega_0 (1+z)^n+\frac{\Omega_0^2}{2\lambda
'}(1+z)^{2n}\right ]^{1/2}
\label{hc}
\end{equation}

Figure \ref{distance} shows the relationship between the
luminosity  distance and the redshift when brane tension parameter
$\lambda '$ and state equation parameter $n$ are taken to various
values respectively. The curves about $\lambda '=10$ show that the
luminosity distance increases faster as the redshift increases
when the state equation parameter $n$ is smaller. From the two
curves when $n=3$ and $\lambda '$ is equal to $10, 100$
respectively, we know how the brane tension parameter affects
the dependence of luminosity distance on the redshift,i.e., when
$\lambda '$ is larger the luminosity increases faster as the
redshift increases.

\section{Conclusions}
In this paper, following the framework of Binetruy et
al.\cite{binetruy}, we investigated kinds of cosmological effects
of the brane Universe without the explicit a cosmological constant-like term. First, 
we deduced the critical density in
the brane Universe, which is different from that in the standard
cosmology and is a function of brane tension parameter. Then we
found a general solution of the brane Universe and discussed the
dependence of the scale factor on the brane tension parameter as
well as the state equation parameter. We discussed the
relationship between the deceleration parameter and the brane
tension parameter, and found that the deceleration is a monotone
decreasing function of the brane tension parameter $\lambda '$. We
also discussed the cosmic time -redshift relation and the
dependence of the age of the brane Universe on the brane tension
parameter. Finally, we discussed the dependence of luminosity
distance on the redshift for different values of the brane tension
parameter $\lambda '$ and the state equation parameter.

We illustrate that when the brane parameter goes infinite large all the results 
approach the standard Friedmann cosmology ones without a cosmological constant-like term.  
The dark energy can be accomodated in the standard Friedmann cosmology based on the 
relativistic gravitational theory, G.R. with a cosmological constant-like term that may 
accelerate our deep understanding to the implying physical world as well as the fundamental
 theories describing it,  and it can also be possiblly interpreted in the extended gravity 
scenarios and cosmological models, which attract our further continnuous research work.

So far brane cosmology is still under frequent discussions with
many theoretical speculations, like the interesting one by using
it trying to describe the Universe accelerating observations.
Surely, we can argue that for the low energy limit the brane
cosmology model can be returned the standard cosmology model. But
 for trying to explore the matching conditions we need more detail
 analysis to the universe classical tests with the brane tension
parameter.

Acknowledgements

The paper has been improved greatly with the comments from communication with Dr. T.Harko, who has provided many instructive advice. The final writtings of this paper is finished in X.H.Meng's academic visit at Hiroshima University supported from JSPS Fellowship. He is very grateful to the theoretical physics group for their extended hospitality to him, especially  Dr.T.Inagaki who provides him lots of helps and makes his stay convenience and enjoyable.
The author has also benefited a lot from discussions with many
people, especially, Profs. X.P.Wu, Y.Q.Yu, D.H.Zhang, X.M.Zhang, Y.Z.Zhang,
 and Y.Zhang. This work is supported by grant of No.
NK-BRSF G19990753  from Ministry of Science and Technology of
China

\end{document}